\documentstyle[twoside,fleqn,amsmath,amsfonts,amssymb,amsthm]{article}
\makeatletter
\def\fileversion{v2.6}
\def\filedate{24 November 1993}

\newdimen\@bls                    % \@b(ase)l(ine)s(kip)
\@bls=\baselineskip               % \@bls ~= \baselineskip for \normalsize
\advance\@bls -1ex                % (fudge term)
\newdimen\@eps                    %
\def\section{\@startsection{section}{1}{\z@}
  {1.5\@bls plus 0.5\@bls}{1\@bls}{\normalsize\bf}}
\def\subsection{\@startsection{subsection}{2}{\z@}
  {1\@bls plus 0.25\@bls}{\@eps}{\normalsize\bf}}
\def\subsubsection{\@startsection{subsubsection}{3}{\z@}
  {1\@bls plus 0.25\@bls}{\@eps}{\normalsize\bf}}
\def\paragraph{\@startsection{paragraph}{4}{\parindent}
  {1\@bls plus 0.25\@bls}{0.5em}{\normalsize\bf}}
\def\subparagraph{\@startsection{subparagraph}{4}{\parindent}
  {1\@bls plus 0.25\@bls}{0.5em}{\normalsize\bf}}

\def\@sect#1#2#3#4#5#6[#7]#8{\ifnum #2>\c@secnumdepth
  \def\@svsec{}\else
  \refstepcounter{#1}\edef\@svsec{\csname the#1\endcsname.\hskip0.5em}\fi
  \@tempskipa #5\relax
  \ifdim \@tempskipa>\z@
    \begingroup
      #6\relax
      \@hangfrom{\hskip #3\relax\@svsec}{\interlinepenalty \@M #8\par}%
    \endgroup
    \csname #1mark\endcsname{#7}\addcontentsline
      {toc}{#1}{\ifnum #2>\c@secnumdepth \else
        \protect\numberline{\csname the#1\endcsname}\fi #7}%
  \else
    \def\@svsechd{#6\hskip #3\@svsec #8\csname #1mark\endcsname
      {#7}\addcontentsline{toc}{#1}{\ifnum #2>\c@secnumdepth \else
        \protect\numberline{\csname the#1\endcsname}\fi #7}}%
  \fi \@xsect{#5}}

\long\def\@makefigurecaption#1#2{\vskip 10mm #1. #2\par}

\long\def\@maketablecaption#1#2{\hbox to \hsize{\parbox[t]{\hsize}
  {#1 \\ #2}}\vskip 0.3ex}

\def\fnum@figure{Figure \thefigure}
\def\figure{\let\@makecaption\@makefigurecaption \@float{figure}}
\@namedef{figure*}{\let\@makecaption\@makefigurecaption \@dblfloat{figure}}

\def\table{\let\@makecaption\@maketablecaption \@float{table}}
\@namedef{table*}{\let\@makecaption\@maketablecaption \@dblfloat{table}}

\textfloatsep=\floatsep           % Space between main text and floats
\intextsep=\floatsep              % Space between in-text figures and
\long\def\@makefntext#1{\parindent 1em\noindent\hbox{${}^{\@thefnmark}$}#1}

\def\maketitle{\begingroup        % Initialize generation of front-matter
    \def\thefootnote{\fnsymbol{footnote}}%
    \newpage \global\@topnum\z@
    \@maketitle \@thanks
  \endgroup
  \let\maketitle\relax \let\@maketitle\relax
  \gdef\@thanks{}\let\thanks\relax
  \gdef\@address{}\gdef\@author{}\gdef\@title{}\let\address\relax}

\def\justify@on{\let\\=\@normalcr
  \leftskip\z@ \@rightskip\z@ \rightskip\@rightskip}

\newbox\fm@box                    % Box to capture front-matter in

\def\@maketitle{%                 % Actual formatting of \maketitle
  \global\setbox\fm@box=\vbox\bgroup
    \vskip 8mm                    % 930715: 8mm white space above title
    \raggedright                  % Front-matter text is ragged right
    \hyphenpenalty\@M             % and is not hyphenated.
    {\Large \@title \par}         % Title set in larger font.
    \vskip\@bls                   % One line of vertical space after title.
    {\normalsize                  % each author set in the normal
     \@author \par}               % typeface size
    \vskip\@bls                   % One line of vertical space after author(s).
    \@address                     % all addresses
  \egroup
  \twocolumn[%                    % Front-matter text is over 2 columns.
    \unvbox\fm@box                % Unwrap contents of front-matter box
    \vskip\@bls                   % add 1 line of vertical space,
    \unvbox\abstract@box          % unwrap contents of abstract boxes,
    \vskip 2pc]}                  % and add 2pc of vertical space

\newcounter{address}
\def\theaddress{\alph{address}}
\def\@makeadmark#1{\hbox{$^{\rm #1}$}}

\def\address#1{\addressmark\begingroup
  \xdef\@tempa{\theaddress}\let\\=\relax
  \def\protect{\noexpand\protect\noexpand}\xdef\@address{\@address
  \protect\addresstext{\@tempa}{#1}}\endgroup}
\def\@address{}

\def\addressmark{\stepcounter{address}%
  \xdef\@tempb{\theaddress}\@makeadmark{\@tempb}}

\def\addresstext#1#2{\leavevmode \begingroup
  \raggedright \hyphenpenalty\@M \@makeadmark{#1}#2\par \endgroup
  \vskip\@bls}

\newbox\abstract@box              % Box to capture abstract in

\def\abstract{%
  \global\setbox\abstract@box=\vbox\bgroup
  \small\rm
  \ignorespaces}
\def\endabstract{\par \egroup}

\def\thebibliography#1{\section*{REFERENCES}\list{\arabic{enumi}.}
  {\settowidth\labelwidth{#1.}\leftmargin=1.67em
   \labelsep\leftmargin \advance\labelsep-\labelwidth
   \itemsep\z@ \parsep\z@
   \usecounter{enumi}}\def\makelabel##1{\rlap{##1}\hss}%
   \def\newblock{\hskip 0.11em plus 0.33em minus -0.07em}
   \sloppy \clubpenalty=4000 \widowpenalty=4000 \sfcode`\.=1000\relax}

\newcount\@tempcntc
\def\@citex[#1]#2{\if@filesw\immediate\write\@auxout{\string\citation{#2}}\fi
  \@tempcnta\z@\@tempcntb\m@ne\def\@citea{}\@cite{\@for\@citeb:=#2\do
    {\@ifundefined
       {b@\@citeb}{\@citeo\@tempcntb\m@ne\@citea
        \def\@citea{,\penalty\@m\ }{\bf ?}\@warning
       {Citation `\@citeb' on page \thepage \space undefined}}%
    {\setbox\z@\hbox{\global\@tempcntc0\csname b@\@citeb\endcsname\relax}%
     \ifnum\@tempcntc=\z@ \@citeo\@tempcntb\m@ne
       \@citea\def\@citea{,\penalty\@m}
       \hbox{\csname b@\@citeb\endcsname}%
     \else
      \advance\@tempcntb\@ne
      \ifnum\@tempcntb=\@tempcntc
      \else\advance\@tempcntb\m@ne\@citeo
      \@tempcnta\@tempcntc\@tempcntb\@tempcntc\fi\fi}}\@citeo}{#1}}

\def\@citeo{\ifnum\@tempcnta>\@tempcntb\else\@citea
  \def\@citea{,\penalty\@m}%
  \ifnum\@tempcnta=\@tempcntb\the\@tempcnta\else
   {\advance\@tempcnta\@ne\ifnum\@tempcnta=\@tempcntb \else
\def\@citea{--}\fi
    \advance\@tempcnta\m@ne\the\@tempcnta\@citea\the\@tempcntb}\fi\fi}

\def\ps@crcplain{\let\@mkboth\@gobbletwo
     \def\@oddhead{\reset@font{\sl\rightmark}\hfil \rm\thepage}%
     \def\@evenhead{\reset@font\rm \thepage\hfil\sl\leftmark}%
     \let\@oddfoot\@empty
     \let\@evenfoot\@oddfoot}

\ps@crcplain                    % modified 'plain' page style

\newif\if@fewtab\@fewtabtrue
\makeatother
\newcommand{\Rn}{{\Bbb R}}

\newcommand{\Zn}{{\Bbb Z}}

\newtheorem*{Thm}{Theorem}

\renewcommand{\[}{\begin{eqnarray}}
\newcommand{\nn}{\nonumber}
\newcommand{\non}{\nonumber \\}

\renewcommand{\]}{\end{eqnarray}}

\newcommand{\een}{\end{enumerate}}
\newcommand{\ben}{\begin{enumerate}}
\newcommand{\ga}{\alpha}

\newcommand{\gd}{\delta}

\newcommand{\gep}{\epsilon}
\newcommand{\gf}{\varphi}
\newcommand{\gl}{\lambda}
\newcommand{\gL}{\Lambda}

\newcommand{\gO}{\Omega}
\newcommand{\gp}{\psi}

\newcommand{\gr}{\rho}

\newcommand{\gx}{\xi}

\newcommand{\gz}{\zeta}
\newcommand{\cl}{\ell}

\newcommand{\cF}{{\cal F}}

\newcommand{\cP}{{\cal P}}
\newcommand{\cPL}{{{\cal P}(\vgL)}}
\newcommand{\cPl}{{{\cal P}^{(\vgl)}}}
\newcommand{\cQ}{{\cal Q}}

\newcommand{\fg}{{\frak g}}
\newcommand{\fh}{{\frak h}}
\newcommand{\fn}{{\frak n}}

\newcommand{\sX}{{\cal X}}

\renewcommand{\vec}[1]{{\boldsymbol{#1}}}
\newcommand{\va}{\vec{a}}

\newcommand{\vk}{\vec{k}}

\newcommand{\vp}{\vec{p}}
\newcommand{\vP}{\vec{P}}
\newcommand{\vq}{\vec{q}}

\newcommand{\vX}{\vec{X}}
\newcommand{\vr}{\vec{r}}

\newcommand{\vs}{\vec{s}}

\newcommand{\vv}{\vec{v}}

\newcommand{\vga}{\vec{\ga}}

\newcommand{\vgd}{\vec{\gd}}
\newcommand{\vgl}{\vec{\gl}}
\newcommand{\vgL}{\vec{\gL}}

\newcommand{\vgr}{\vec{\gr}}

\newcommand{\vgx}{\vec{\gx}}

\newcommand{\vo}{\vec{0}}
\newcommand{\Oint}{\oint\limits}
\newcommand{\res}[1]{\oint\!\frac{{\rm d}#1}{2\pi i}\,}
\newcommand{\Res}[2]{\Oint_{#1}\!\frac{{\rm d}#2}{2\pi i}\,}

\newcommand{\Vir}{{\rm Vir}}

\newcommand{\mult}{{\rm mult}}

\newcommand{\re}{{\rm e}}
\newcommand{\frc}[2]{{\textstyle \frac{#1}{#2}}}

\renewcommand{\|}{\,|\,}
\renewcommand{\.}{\cdot}
\newcommand{\X}{\!\cdot\!}

\newcommand{\ket}[1]{{|#1\rangle}}
\newcommand{\pord}[1]{\mbox{\large\bf:} #1 \mbox{\large\bf:}}
\newcommand{\xord}[1]{{}_\times^\times #1 {}_\times^\times}
\newcommand{\cord}[1]{{}_\circ^\circ #1 {}_\circ^\circ}

\newcommand{\hv}{{h^\vee}}
\newcommand{\bfg}{{\bar{\fg}}}

\newcommand{\bfh}{{\bar{\fh}}}

\newcommand{\bgD}{{\bar{\Delta}}}
\newcommand{\bQ}{{\bar{Q}}}

\newcommand{\vkl}{{\vk_\cl}}

\newcommand{\Ai}[1]{{A^i_{#1}}}

\newcommand{\Er}[1]{{E^{\vr}_{#1}}}
\newcommand{\Emr}[1]{{E^{-\vr}_{#1}}}
\newcommand{\Es}[1]{{E^{\vs}_{#1}}}
\newcommand{\Erb}[1]{{E^{\vr+\vs}_{#1}}}
\newcommand{\XI}{\sX^i}
\newcommand{\vcX}{\vec{\sX}}
\newcommand{\vcQ}{\vec{\cQ}}

\newcommand{\sL}[1]{{{\cal L}_{#1}}}

\newcommand{\tr}[1]{t^{[\cl]}_{#1}}
\title{\vbox{\vskip-48pt\hbox to\hsize{\hfill \normalsize \tt
             hep-th/9612179}\vskip54pt}
       Beyond the Frenkel--Kac--Segal construction of affine Lie
       algebras\thanks{Contribution to Proceedings of the 30th Int.\
       Symposium Ahrenshoop on the Theory of Elementary Particles,
       Buckow, Germany, August 27-31, 1996}}
\author{R.~W.~Gebert%
        \address{Institute for Advanced Study,
                 School of Natural Sciences,\\
                 Princeton, NJ 08540, U.S.A.}%
        \thanks{Supported by {\em Deutsche Forschungsgemeinschaft} under
       Contract DFG Ge 963/1-1.}}

\begin{document}
\thispagestyle{empty}
\begin{abstract}
This contribution reviews recent progress in constructing affine Lie
algebras at arbitrary level in terms of vertex operators. The string
model describes a completely compactified subcritical chiral bosonic
string whose momentum lattice is taken to be the (Lorentzian) affine
weight lattice. The main feature of the new realization is the
replacement of the ordinary string oscillators by physical DDF
operators, whereas the unphysical position operators are substituted
by certain linear combinations of the Lorentz generators. As a side
result we obtain simple expressions for the affine Weyl translations
as Lorentz boosts. Various applications of the construction are
discussed.
\end{abstract}

\maketitle

\section{MOTIVATION}
Recent developments \cite{HarMoo96,HarMoo97,DiVeVe96} indicate that
finite and affine Lie algebras are not sufficient for the description of
symmetries in string theory. It is quite likely that hyperbolic
Kac--Moody algebras and their (super)extensions to Borcherds
(super)algebras, also called generalized Kac--Moody (super)algebras,
will play an important role for the understanding of certain string
symmetries.

Unfortunately, from the mathematical point of view not much is known
about these infinite-dimensional Lie algebras. One could also turn
this into an advantage by interpreting it as a promising sign since
history has taught us that new fundamental developments in physics
very often involve new mathematics.

At the physical side, there is a class of certain (completely
toroidally compactified) string models where hyperbolic and Lorentzian
Kac--Moody algebras and some of their Borcherds extensions are
explicitly realized. More specifically, they arise as Lie algebras of
physical string states (see e.g.\ the review \cite{Gebe93}) with Lie
bracket given by $[\gp,\gf]:= \res{z}V(\gp,z)\gf$ for physical states
$\gp,\gf$ and associated vertex operator $V(\ ,z)$. At first sight it
may seem rather awkward that the physical states themselves form a Lie
algebra. Nonetheless, the bracket has a simple interpretation in terms
of tree level scattering of physical string states \cite{GeNiWe96}.

The general feature of hyperbolic Kac--Moody algebras is that they can
be decomposed into an infinite direct sum of highest (and of lowest)
weight modules for the underlying affine subalgebra. In the above
string realization this entails that one has to deal with infinitely
many affine highest weight modules within a single physical state
space. To handle this problem one would need a physical string vertex
operator construction of affine Lie algebras at arbitrary level. This
contribution reports on recent progress \cite{GebNic97} in this
question which was obtained in collaboration with H.~Nicolai.

\section{COMPACTIFIED BOSONIC STRING}
We consider a (chiral half of a) closed bosonic string moving on a
$d$-dim Minkowskian torus as spacetime. Uniqueness of the quantum
mechanical wave function then forces the center of mass momenta of the
string to lie on an even Lorentzian lattice $\gL$. From a
phenomenological point of view it is certainly not very plausible to
compactify all spacetime dimensions since we know that there should be
at least four macroscopic dimensions. As mentioned above, the
philosophy behind our approach is that in such models new
infinite-dimensional Lie algebras occur as symmetries which might
prove useful in different contexts. After all, these models provide
the only realizations we know so far.

The Fock space $\cF$ is spanned by states of the form
\[ \ga^{\mu_1}_{-m_1}\cdots\ga^{\mu_M}_{-m_M}\ket\vgl, \nn \]
where the $\ga^\mu_m$'s form a $d$-fold oscillator algebra,
\[ [\ga^\mu_m,\ga^\nu_n]=m\eta^{\mu\nu}\gd_{m+n,0}\ , \nn \]
and the groundstates $\ket\vgl=\re^{i\vgl\.\vq}\ket\vo$ for
$\vgl\in\gL$ satisfy ($[q^\mu,p^\nu]=i\eta^{\mu\nu}$,
$p^\mu\equiv\ga^\mu_0$)
\[ p^\mu\ket\vgl=\gl^\mu\ket\vgl,\qquad
   \ga^\mu_m\ket\vgl=0\quad\forall m>0\ . \nn \]

In order to define physical states one has to implement the Virasoro
constraints (with central charge $c=d$)
\[ L_n=\frac12\sum_{m\in\Zn}\pord{\vga_m\X\vga_{n-m}}\,, \nn \]
where $\pord\ldots$ denotes normal-ordering with respect to the string
oscillator modes. Due to the anomaly, however, we proceed \`a la
Gupta--Bleuler in electromagnetism which means that we can impose only
half of the constraints. Hence physical states are conformal primary
states of weight 1,
\[ \cP := \bigoplus_{\vgl\in Q^*}\cPl, \nn \]
where
\[ \cPl := \{\gp\|
   L_n\gp=\gd_{n0}\gp\ \forall n\ge0,\
   p^\mu\gp=\gl^\mu\gp\}. \nn \]
The simplest examples of physical states are tachyons $\ket\va$ with
$\va\in\gL$ and $\va^2=2$, photons $\vgx\X\vga_{-1}\ket\vk$ with
$\vgx\in\Rn^{d-1,1}$, $\vk\in\gL$ and $\vgx\X\vk=\vk^2=0$, etc..

\section{DDF CONSTRUCTION}
It would be nice to have an explicit description of the space of
physical states in the same way as the Fock space is built as an
infinite sum of Heisenberg modules. Such a description is indeed
possible and is provided by the so-called DDF construction
\cite{DeDiFu72,Brow72} adjusted to the discrete model
\cite{GebNic95}. For a given physical momentum $\vgl\in\gL$,
$\vgl^2\le2$, one first has to find a tachyon $\ket\va$ and a
lightlike vector $\vk$ such that $\va\X\vk=1$ and
\[ \vgl=\va-n\vk \qquad\text{with } n:=1-\frc12\vgl^2\ . \nn \]
Such a pair $(\va,\vk)$ always exists as long as we do not require the
vectors to lie on the lattice. We refer to it as a DDF decomposition
of $\vgl$. Next we choose $d-2$ orthonormal polarization vectors
$\vgx^i(\va,\vk)\in\Rn^{d-1,1}$ satisfying $\vgx^i\X\va= \vgx^i\X\vk=
0$, and define the transversal DDF operators
\[ \Ai{m}=\Ai{m}(\va,\vk)
    := \Res0{z}\vgx^i\X\vP(z)\,
                \re^{im\vk\.\vX(z)}, \nn \]
in terms of the Fubini--Veneziano fields
\[ X^\mu(z) &:=&
   q^\mu-ip^\mu\ln z+i\sum_{m\ne0}\frac1m\ga^\mu_mz^{-m}, \non
   P^\mu(z) &:=& i\frac{d}{dz}X^\mu(z)
              = \sum_{m\in\Zn}\ga^\mu_mz^{-m-1}. \nn \]
Using operator product techniques it is straightforward to show that
the $A^i_m$'s obey a $(d-2)$-fold ``transversal'' oscillator algebra,
\[ [A^i_m,A^j_n]=m\gd^{ij}\gd_{m+n,0}. \nn \]

There are also longitudinal DDF operators $A^-_m=A^-_m(\va,\vk)$ whose
complicated expressions are not needed here. They form a
``longitudinal'' Virasoro algebra with central charge $c=26-d$ and
commute with the transversal DDF operators.

The nice thing about the DDF operators is that they constitute a
spectrum-generating algebra for the string; for it can be shown that
they commute with the Virasoro constraints $L_n$ (and hence map
physical states into physical states) and $\cPl$ is spanned by
\[ A^{i_1}_{-n_1}\cdots A^{i_N}_{-n_N}
   A^-_{-m_1}\cdots A^-_{-m_M}\ket\va\ , \nn \]
where $n_1+\ldots+m_M=1-\frc12\vgl^2$. Note that the DDF operators
also account for an explicit description of the null physical states
contained in $\cPl$ due to the relation $A^-_{-1}\ket\va\propto
L_{-1}\ket{\va-\vk}$, which can be verified by a straightforward
computation.

\section{AFFINE LIE ALGEBRA}
Let $\bfg$ be a finite-dimensional simple Lie algebra of type $ADE$
and rank $d-2$ ($d\ge3$). We denote the root system of $\bfg$ by
$\bgD$. Consider the associated nontwisted affine Lie algebra
$\fg=\fn_+\oplus\fh\oplus\fn_-$ (see e.g.\ \cite{Kac90,MooPia95}). The
$d$-dimensional affine Cartan subalgebra $\fh$ then contains a central
element and a scaling element denoted by $K$ and $d$,
respectively. The corresponding elements in $\fh^*$ are called null
root $\vgd$ and basic fundamental weight $\vgL_0$, respectively.  We
have the scalar products $\vgd\X\vgL_0=1$ and $\vgd^2= \vgL_0^2=
\vgd\X\vr_i= \vgL_0\X\vr_i= 0$ for the real simple roots $\vr_i$
($1\le i\le d-2$) of $\bfg$.

In the above string model we shall now choose for the momentum lattice
$\gL$ the affine weight lattice $Q^*$, which is even Lorentzian as
required.

Recall that an irreducible level-$\cl$ highest weight module $L(\vgL)$
for an affine Lie algebra $\fg$ is determined by the following data: a
vacuum vector $v_{\vgL}$, a dominant integral weight $\vgL\in Q^*$,
and a weight system $\gO(\vgL)$ with appropriate weight
multiplicities. Without loss of generality we may assume that
$\vgL^2=2$, because $\vgL+z\vgd$ for any $z\in\Rn$ gives rise to an
isomorphic $\fg$-module.

An affine Cartan--Weyl basis in terms of integrated vertex operators
can be introduced as follows. Let $\vgx^i$ be a set of $d-2$
orthonormal polarization vectors associated with the DDF decomposition
$(\vgL,\vkl)$ where $\vkl:= \frc1\cl\vgd$. On
\[ \cPL:=\bigoplus_{\vgl\in\gO(\vgL)}\cPl, \nn \]
which is a subspace of the space of physical string states, $\cP$, we
define
\begin{subequations} \label{CW-basis}
\[ K &:=& \vgd\X\vp, \qquad
   d := \vgL_0\X\vp, \\[1ex]
   H^i_m &:=&
     \Res0{z}\vgx^i\X\vP(z)\,\re^{im\vgd\.\vX(z)}, \\[1ex]
   \Er{m} &:=&
     \Res0{z}\pord{\re^{i(\vr+m\vgd)\.\vX(z)}}c_{\vr}, \label{step} \]
   \end{subequations}
for all $\vr\in\bgD$, where $c_{\vr}$ denotes some appropriate cocycle
factor satisfying $c_{\vr}\ket\vs= \gep(\vr,\vs)\ket\vs$ for some
2-cocycle $\gep$. One can show \cite{Fren85,GodOli85} that these
operators obey the commutation relations
\[ [H^i_m,H^j_n]
   &=& \cl m\gd^{ij}\gd_{m+n,0} , \nn \\[.5ex]
   [H^i_m,\Er{n}]
   &=& (\vgx^i\X\vr)\Er{m+n}, \nn \\[.5ex]
   [\Er{m},\Es{n}]
   &=& \begin{cases}
       0 & \text{if $\vr\X\vs\ge0$}, \nn \\
       \gep(\vr,\vs)\Erb{m+n} & \text{if $\vr\X\vs=-1$}, \nn \\
       H^{\vr}_{m+n}+\cl m\gd_{m+n,0}  & \text{if $\vr\X\vs=-2$},
       \end{cases} \nn \\[.5ex]
   [d,H^i_m]
   &=& mH^i_m, \qquad
   [d,\Er{m}]
   = m\Er{m}, \nn \\[.5ex]
   [K,x]
   &=& 0 \qquad\forall x\in\fg.\nn  \]
This gives a level-$\cl$ vertex operator realization of $\fg$ on
$\cPL$. In fact, we can identify the vacuum vector $v_{\vgL}$ in
$L(\vgL)$ with the tachyonic groundstate $\ket{\vgL}$ in $\cPL$ to
conclude that\footnote{An especially nice feature of this string realization
   is the fact that both the vacuum vector and the null vector
   conditions in $L(\vgL)$ immediately follow from momentum
   conservation and the physical state condition $L_0\gp=\gp$.}
\[  L(\vgL)\hookrightarrow\cPL,\qquad
    L(\vgL)_{\vgl}\hookrightarrow\cPl. \nn \]

Let us make some remarks.

We observe that the $H^i_m$'s, which make up the homogeneous
Heisenberg subalgebra of $\fg$, are nothing but the transversal DDF
operators $A^i_{\cl m}$ and thus play the role of spectrum-generating
elements. Only for $\cl=1$ the Heisenberg subalgebra and the
transversal oscillator algebra are identical. One may wonder why there
is only one set of polarization vectors $\vgx^i(\vgL,\vkl)$ and DDF
operators $\Ai{m}(\vgL,\vkl)$ although one has such data for each
$\vgl\in\gO(\vgL)$. One easily shows, however, that the polarization
vectors can always be chosen such that they differ only by vectors
proportional to $\vgd$ when going from one $\vgl$ to another. But
since $\vgd\X\vP(z)\,\re^{im\vgd\.\vX(z)}$ for $m\ne0$ is a total
derivative, we conclude that the operators $H^i_m$ are indeed universally
defined on $\cPL$.

Below, we will see that only transversal physical states can occur in
the affine highest weight module $L(\vgL)$. Hence we effectively deal
with the embedding $L(\vgL)_{\vgl}\hookrightarrow {\cal
P}^{(\vgl)}_{\rm transv.}$ and have the following universal estimate
for affine weight multiplicities at arbitrary level:
\[ \mult_{L(\vgL)}(\vgl)
    \le \dim {\cal P}^{(\vgl)}_{\rm transv.}=p_{d-2}(1-\frc12\vgl^2), \nn \]
where $p_{d-2}(n)$ counts the partition of $n$ into ``parts'' of $d-2$
``colours''.

\section{RESULT}
In view of the fact that the operators of the Cartan--Weyl basis
\eqref{CW-basis} are integrated vertex operators associated with
physical string states and thus are physical operators by
construction, it is sensible to ask whether they may be directly
expressed in terms of the DDF operators, i.e., in a manifestly
physical form. Note that this is already the case for the $H^i_m$'s
(see above remark). So only the step operators \eqref{step} remain to
be dealt with. It is certainly true that given a step operator acting
on some physical state, the resulting physical state can be written in
the DDF basis. So the precise question is whether there is some
unifying formula (independent of the state acted on) for the step
operators in terms of physical operators. Note that this is a highly
nontrivial problem because of the exponential dependence of both the
step operators and the DDF operators on the string oscillators. The
main result of \cite{GebNic97} gives an affirmative answer to the
above question.
\begin{Thm}
On $\cPL$, one can rewrite the step operators as follows:
\[ \Er{m}\big|_{\cPL} = \Res0{z} z^{\cl m}
  \xord{\re^{i\vr\.\vcX(z)}}c_{\vr}, \label{stepnew} \]
where
\[ \XI{}(z)
    &:=& \cQ^i- i\Ai{0}\ln z+i\sum_{m\ne0}\frac1m\Ai{m}z^{-m}, \non
   { \cQ^i}
    &:=& (\vgx^i)_\mu(\vkl)_\nu M^{\mu\nu}, \nn \]
with Lorentz generators
\[ { M^{\mu \nu}}:= q^\mu p^\nu - q^\nu p^\mu - i\sum_{n\neq 0}
   \frac{1}{n} \ga_{-n}^{[\mu} \ga_n^{\nu ]}, \nn \]
and $\xord\ldots$ denotes normal-ordering with respect to the mode
indices of the transversal DDF operators.
\end{Thm}

Let us discuss some aspects of this formula.

The above vertex operator construction may be characterized as
``doubly transcendental'' due to the appearance of DDF operators in
the exponential. Furthermore, it is ``purely transversal'' since only
transversal DDF operators are involved.

At first sight it is surprising that the Lorentz generators pop up. On
the other hand, it is clear that the unphysical position operators
$q^\mu$ in the step operators $\Er{m}$, which generate a momentum
shift by $r^\mu$, must be replaced by some physical operators other
than the DDF operators, which shift the momentum only along the
direction of the affine null root. Therefore one is inevitably led to
consider the Lorentz generators which are physical and which rotate
the momentum vectors. If we introduce momentum operators $\cP^i\equiv
\Ai{0}$ then the Lorentz generators $\cQ^i$, which replace the
$q^i$'s, are canonically conjugate to them and may be regarded as
``physical position operators'', viz.\ $[\cQ^i,\cP^j]= i\gd^{ij}$.

As a side result, the new formula for the step operators provides a
natural interpretation of affine Weyl translations as Lorentz
boosts. One finds that
\[ \re^{i\vr\.\vcQ}\vv\X\vga_m\re^{-i\vr\.\vcQ}
   =\tr{\vr}(\vv)\X\vga_m \nn \]
for all $\vv\in\fh^*$, $\vr\in\bQ$ (finite root lattice), where
\[  \tr{\vr}(\vv) := \vv + (\vv\X\vkl)\vr -
    \left[(\vv\X\vkl)\frc12\vr^2 + \vr\X\vv\right]\vkl. \nn \]

Indeed, the last expression is precisely the formula for an affine
Weyl translation (see e.g.\ \cite{Kac90}) which arises in the
decomposition of the affine Weyl group into a semidirect product of
the finite Weyl group and the affine translation group isomorphic to
the finite root lattice. Clearly, the affine null root is invariant
under the translations $\tr{\vr}$. In this way the affine Weyl group
becomes a discrete subgroup of ${\rm ISO}(d-2)$, the subgroup of the
full Lorentz group ${\rm SO}(d-1,1)$ leaving fixed a given lightlike
vector. We can thus think of the affine Weyl group as a ``dimensional
null reduction'' of the full Lorentzian Weyl group.

Formula \eqref{stepnew} for the step operators resembles very much the
famous Frenkel--Kac--Segal construction of level-1 affine Lie algebras
\cite{FreKac80,Sega81}. There, the momentum lattice $\gL$ is taken to
be the Euclidean finite root lattice $\bQ$ and the step operators are
the (generically) unphysical Laurent modes of the tachyon vertex
operators: $\Er{m} := \res{z} z^m \pord{\re^{i\vr\.\vX(z)}}c_{\vr}$, in
terms of the original Fubini--Veneziano field $\vX$. Therefore the
theorem provides a genuine generalization of the Frenkel--Kac--Segal
construction to arbitary level. Furthermore, the DDF operators are
promoted to physical string oscillators similar to the string
oscillators appearing in the light cone gauge formulation of string
theory.

Finally, in view of the last remark, we would like to mention an
additional feature of the new construction. So far, the discussion was
limited to the space $\cPL$ associated with any dominant integral
affine weight $\vgL$. But the whole space of physical string states,
$\cP$, comprises all such representation spaces $\cPL$. Consequently,
the new vertex operator construction allows for a simultaneous
treatment of infinitely many affine highest weight representations at
arbitrary level within a single state space, which is especially
relevant for the understanding of hyperbolic Kac--Moody
algebras. Since these modules are purely transversal it is almost
obvious which role the longitudinal DDF operators will play in this
framework. Namely, they must map between different $\fg$-modules and
hence act as (often level-changing) intertwining operators (see
\cite{GebNic97} for a discussion of these issues).

\section{SUGAWARA OPERATORS}
Given an affine Lie algebra $\fg$ with Cartan--Weyl basis $H^i_n$,
$\Er{n}$ ($1\le i\le d-2$, $\vr\in\bgD$), it is well-known that there
is a Virasoro algebra $\Vir_{\fg}$ associated with it such that any
representation of $\fg$ can be extended to a representation of the
semidirect product $\fg{\Bbb o}\Vir_{\fg}$. Indeed, the so-called
Sugawara operators
\[ \sL{m}
   &:=& \frac1{2(\cl+\hv)}\sum_{n\in\Zn}
        \bigg[\sum_{i=1}^{d-2}\cord{H^i_nH^i_{m-n}}+ \non
   & & \phantom{2(\cl+\hv)\sum\bigg[}
        +\sum_{\vr\in\bgD}\cord{\Er{n}\Emr{m-n}}\bigg] \label{sug} \]
form a Virasoro algebra with central charge $c_\cl:=
\frac{\cl\dim\bfg}{\cl+\hv}$, where $\cl$ and $\hv$ denote the level
and the dual Coxeter number, respectively, and the symbol
$\cord\ldots$ refers to normal-ordering with respect to the mode
indices of the affine generators. If we insert the expression
\eqref{stepnew} for the step operators, we arrive at the following new
formula for the Sugawara operators at arbitrary level in terms of
transversal DDF operators \cite{GeKoNi96}:
\[ \sL{m}
   &=&\frac1{2\cl}\sum_{n\in\Zn}
              \sum_{i=1}^{d-2}\xord{\Ai{\cl n}\Ai{\cl(m-n)}} \non
   & &{}+\frac{\hv}{2\cl(\cl+\hv)}\sum_{n\neq 0(\cl)}
              \sum_{i=1}^{d-2}\xord{\Ai{n}\Ai{\cl m-n}} \non
   & &{}+\frac{(\cl^2-1)(d-2)\hv}{24\cl(\cl+\hv)}\gd_{m,0} \non
   & &{}-\frac1{2\cl(\cl+\hv)}\sum_{\vr\in\bgD}
              \sum_{p=1}^{\cl-1}\frac1{|\gz^p-1|^2}\times \non
   & &\qquad\times
              \Oint\!\frac{dz}{2\pi i}\,z^{\cl m -1}
              \xord{\re^{i\vr\.\left[\vcX(\gz^pz)-\vcX(z)\right]}},
\label{sugnew} \]
where $\gz:=\re^{2\pi i/\cl}$.

Only special cases of this formula had been known before. For level
$\cl=1$ one easily finds that
\[ \sL{m}=\frac12\sum_{n\in\Zn}\sum_{i=1}^{d-2}\xord{H^i_nH^i_{m-n}},
\nn \]
which is referred to in the literature as the equivalence of the
Virasoro and the Sugawara construction. At arbitrary level, the action
of the zero mode operator on an affine highest weight vector
$\ket\vgL$ is given by the simple formula
\[ \sL0\ket\vgL=\frac{(\bar{\vgL}+2\bar{\vgr})\X\bar{\vgL}}%
                       {2(\cl+\hv)}\ket\vgL,\nn \]
where $\bar{\vgr}$ denotes the finite Weyl vector and $\bar{\vgL}$ is
the projection of $\vgL$ onto $\bfh$. Previously, one had to relie on
properties of the affine Casimir operator in order to derive this
result. With the new formula \eqref{sugnew} at hand, the above two
expressions can be immediately read off (for the second expression one
also has to invoke some simple identity for sums over roots of unity).

A remarkable feature of the last term in formula \eqref{sugnew} is its
nonlocality. Responsible for this is the factor $z^{\cl m}$ in
\eqref{stepnew}. Indeed, in evaluating the expression
$\sum_{n\in\Zn}\cord{\Er{n}\Emr{m-n}}$ this leads to a factor of
$\sum_{n\ge0}z^{-\cl n}w^{\cl(m+n)}= \frac{z^\cl w^{\cl
m}}{z^\cl-w^\cl}$ in the corresponding operator product expansion.
Consequently, we pick up additional poles at $z=\re^{2\pi ip/\cl}w$
($1\le p\le\cl$) when we perform the contour integral in the
$z$-plane. An interpretation of this is that we are not working on the
Riemann sphere but rather on a $\cl$-sheeted covering of it.\footnote{
I am grateful to A.~M.~Semikhatov for pointing this out to me.}

%\bibliographystyle{wsci}
%\bibliography{Liste}

\begin{thebibliography}{10}

\bibitem{HarMoo96}
J.A.~Harvey and G.~Moore, Nucl.\ Phys.\ B463 (1996) 315.

\bibitem{HarMoo97}
J.A.~Harvey and G.~Moore, preprint hep-th/9609017.

\bibitem{DiVeVe96}
R.~Dijkgraaf, E.~Verlinde and H.~Verlinde, preprint hep-th/9607026.

\bibitem{Gebe93}
R.W.~Gebert, Int.\ J.\ Mod.\ Phys.\ A8 (1993) 5441.

\bibitem{GeNiWe96}
R.W.~Gebert, H.~Nicolai and P.C.~West, Int.\ J.\  Mod.\  Phys.\ A11
(1996) 429.

\bibitem{GebNic97}
R.W.~Gebert and H.~Nicolai, preprint DESY 96-166, hep-th/9608014.

\bibitem{DeDiFu72}
E.~{Del Giudice}, P.~{Di Vecchia} and S.~Fubini, Ann.\ Physics 70
(1972) 378.

\bibitem{Brow72}
R.C.~Brower, Phys.\ Rev.\ D6 (1972) 1655.

\bibitem{GebNic95}
R.W.~Gebert and H.~Nicolai, Commun.\ Math.\ Phys.\ 172 (1995) 571.

\bibitem{Kac90}
V.~Kac, Infinite dimensional Lie algebras, Cambridge Univ.\ Press,
Cambridge, 1990.

\bibitem{MooPia95}
R.V.~Moody and A.~Pianzola, Lie Algebras With Triangular
Decomposition, John Wiley \& Sons, New York, 1995.

\bibitem{Fren85}
I.B.~Frenkel, in: Applications of Group Theory in Theoretical Physics,
  pp.\ 325, American Mathematical Society, Providence, 1985.

\bibitem{GodOli85}
P.~Goddard and D.~Olive, in: Vertex Operators in Mathematics and
Physics, J.~Lepowsky, S.~Mandelstam and I.M.~Singer (eds.), pp.\ 51,
Springer, New York, 1985.

\bibitem{FreKac80}
I.B.~Frenkel and V.G.~Kac, Invent.\ Math.\ 62 (1980) 23.

\bibitem{Sega81}
G.~Segal, Commun.\ Math.\ Phys.\ 80 (1981) 301.

\bibitem{GeKoNi96}
R.W.~Gebert, K.~Koepsell and H.~Nicolai, preprint DESY 96-072,
hep-th/9604155, to appear in Commun.\ Math.\ Phys.\ (1997).

\end{thebibliography}

%
\end{document}